\documentstyle[aps,preprint,epsfig]{revtex}
\begin{document}
\draft              
\title{Instanton picture of the spin tunneling in the Lipkin model}
\author{V. I. Belinicher} 
\address{University of Coimbra, 3000, Coimbra, Portugal \\  and
Institute of Semiconductor Physics, 630090, Novosibirsk, Russia}
\author{C. Providencia, and J. da Providencia}
\address{University of Coimbra, 3000, Coimbra, Portugal}
\date{\today}
\maketitle
\begin{abstract}
  A consistent theory of the ground state energy and its splitting due to 
the process of tunneling for the Lipkin model is presented. For the
functional integral in terms of the spin coherent states for the 
partition function of the model we accurately
calculate the trivial and the instanton saddle point contributions.
We show that such calculation has to be perfomed very accurately
taking into account the discrete nature of the functional integral.
Such accurate consideration leads to finite corrections to a naive continous
consideration. We present comparison with numerical calculation of the 
ground state energy and the tunneling splitting and with the results obtained
by the quasiclassical method and get excellent agreement.      
\vskip 0.5cm 
\noindent 
\end{abstract}
\narrowtext
\section{Introduction}   \ \
\label{sec:Int}
  Lipkin, Meshkov and Glick \cite{Lip} proposed in 1965 
an exactly solvable 
two-level many-fermion
model which has been used to test various kinds of 
many-body theories.
The model Hamiltonian reads
\begin{eqnarray}  \label{a}
H=\tilde{\omega}\hat{S}_0-\frac{1}{2}f(\hat{S}_+^2+\hat{S}_-^2)
\end{eqnarray}
where $f$ is the coupling constant, $\tilde{\omega}$ is 
the energy
difference between the two levels,
\begin{eqnarray}  \label{b}
&&\hat{S}_0=\frac{1}{2}\sum_{m=1}^\Omega(c_{+m}^{\dag}c_{+
m}-
c_{-m}^{\dag}c_{-m}),
\\ \nonumber
&&\hat{S}_+={1\over 
2}\sum_{m=1}^\Omega
c_{+m}^{\dag}c_{-m},\quad \hat{S}_-=\hat{S}_+^{\dag},
\end{eqnarray} 
and $\Omega$ is the degree of degeneracy of each level.
The single particle states are labeled by the quantum 
numbers $\pm m.$
The operators $\hat{S}_0, \hat{S}_\pm$ satisfy the 
commutation relations of
the $su(2)$ algebra. 

The transformation associated with 
the unitary operator
${\rm e}^{i\pi\hat{S}_0}$       
leaves the Hamiltonian invariant. If $\chi 
=f\Omega/\tilde{\omega}<1,$ the mean field
groundstate of the model is classified according to the  
trivial representation
of the symmetry group. The symmetry is 
spontaneously broken if
$\chi>1.$ 

A splitting of the levels is then observed, 
which is
analogous to the occurrence of rotational bands in 
deformed nuclei.
The model is also of great interest in condensed matter 
physics, since it is
related to the anysotropic Heisenberg model. In the 
strong coupling limit,
$\chi\gg 1,$ the level splitting vanishes for odd 
$\Omega.$ This behaviour is
easily understood in the framework of Kramers theorem 
and is related to the
well known phenomenon of tunnelling suppression for odd 
spin anysotropic
Heisenberg ferromagnets. 

In this paper we formulate a general physical picture of the calculation of
the ground state energy and its splitting by the instanton method. 
Such method was applied to the anisotropic Heisenberg model for description
of the tunneling of the magnetic moment of small magnetic particles with large
spin \cite{Enz,Chu1,Chu2,Los,Del,Gal}. Although the physical picture
of the spin tunneling was formulated in these papers the quantitative results
of the papers \cite{Chu1,Chu2,Los,Del,Gal} are not completely correct.

 We show that special care is required when computing the instanton contributions in 
order to take also correctly into account the small amplitude quantum 
fluctuations. On more technical language the functional determinants have to be
calculated very accurately taking properly into account 
the operator ordering generating
the functional integral for the partition sum. We show that it is essential to
keep in mind that the fuctional integral is generated by the spin coherent
state method. Taking accurately into account the discrete nature in time 
of the functional leads to essential corrections to the ground state energy and
to its tunneling splitting. After taking into account all these contribution
our result completely coinsides with the one of \cite{Enz} obtained by 
the semiclassical
method and the contradiction existing in literature is resolved.

The structure of the paper is as follws. In Sec. \ref{sec:Fun} we present the
functional integral for the Lipkin model. In Sec. \ref{sec:Nai} we review the
the simple results for the tunneling in the framework of the continuous approximation
for the functional integral. In Sec. \ref{sec:Gro} we calculated accurately the
contribution of the trivial saddle point to the partition function and
determine the ground state energy. In Sec. \ref{sec:Mea} we calculeted
the determinant and the measure of integration over time in the framework of
the canonical fromulation of the functional integral. In Sec. \ref{sec:Det}
we calculate accurately the functional determinant for the instanton saddle
point taking into account the corrections due to discretization. In Sec.
\ref{sec:Num} we present comparison between the analytical theory and the
numerical calculations for the Lipkin model and find an excellent agreement 
when additional contributions discovered in this paper are taken into account.

\section {Functional integral for the Lipkin model}\label{sec:Fun} \ \

It is convenient to rewrite 
the Hamiltonian of the model in the rotated reference 
frame 
\begin{eqnarray}  \label{1}
&&\hat{H} = \hat{H}_{2} + \hat{H}_{1},\ \ \ \hat{H}_{1} 
= \tilde{\omega} 
\hat{S}_{y}, \\ \nonumber
&&\hat{H}_{2} = f(\hat{S}_z^2-\hat{S}_x^2) =
 f\biggl({3\over 2}\hat{S}_z^2-{1\over 2}s(s+1)-{1\over 
4}
(\hat{S}_+^2+\hat{S}_-^2)\biggr).
\end{eqnarray}
Here we assume that the magnitude of  spin $s \gg 1$.
 The first term in the Hamiltonian (\ref{1}) describes 
interaction
between nucleons, the second one represents an energy 
splitting  and can be
regarded as an external magnetic field. The choice of 
the coordinate axis is
determined by the convenience condition for further 
calculations.

 The Lipkin model can be reduced to the anisotropic 
Heisenberg model
\cite{Chu1}. For that we add to the Hamiltonian 
(\ref{1}) 
the term $f(S_{x}^2 + S_{y}^2 + S_{z}^2 - s(s+1)) = 0$ 
and  the Hamiltonian (\ref{1})
takes the form
\begin{eqnarray}  \label{2}
\hat{H} = - fs(s+1) + f(2\hat{S}_z^2 + \hat{S}_y^2) + 
\tilde{\omega} S_{y}.
\end{eqnarray} 
In the language of the Heisenberg model the magnetic 
field here is in the direction
of the second axis $y$. The parameter $\lambda$ of the 
paper \cite{Chu1} is
$1/2$ for our case. We concentrate our attention on the 
accurate treatment of the ground
state energy of the model and its tunneling splitting  
when the second term in (\ref{1}) 
is absent or $\tilde{\omega} =0$.

 In this special case the Lipkin model is invariant under 
the time
reflection. In the case of half-integer spin $s$ this 
symmetry leads to the two-fold
Kramer's degeneracy of the ground state \cite{Los,Del}. 
In the case of integer spin $s$ instead
of degeneracy we have the splitting of the ground state. 
This splitting has to be
small because for large $s$ the difference between 
integer and half-integer spin
has to be small. In fact this splitting is exponentially 
small.

 To solve this problem we have used the instanton method 
\cite{Col} applied to
the functional integral for the partition function of 
our spin system in terms of spin 
coherent states \cite{Coh}. The coherent states are 
given by the following formula
\cite{Coh}:
\begin{eqnarray}  \label{3}
&&|z\rangle = (1 + |z|^2)^{-s}\exp(z \hat{S}_+)
|0\rangle,\ \ \ \hat{S}_-|0\rangle=0,
\\ \nonumber
&&\hat{S}_z|0\rangle=-s|0\rangle,
\end{eqnarray} 
where $z$ is a complex number. They possess many 
remarkable properties and with the
help of them the partition function can be represented 
in the form \cite{Coh}:
\begin{eqnarray}  \label{4}
&&Z = Tr[\exp(-T\hat{H})], \\ \nonumber
&&Z = \int_{-\infty}^{\infty} \cdot \cdot \cdot 
\int_{-\infty}^{\infty}
\prod_{n = 0}^{N-1} \frac{(2s+1)dz_n dz^*_n}{\pi (1 + 
|z_n|^2)^2} \exp(A(z)).
\end{eqnarray}
Here the interval of the imaginary time $T$ is split 
into $N$ parts: 
$T = N\Delta$, and in every section an integration over 
$z'_n = \mbox{Re}z_n$ and
$z''_n = \mbox{Im}z_n$ is performed. The action of the system $A(z)$ has 
the form
\begin{eqnarray}  \label{5}
&&A(z) = \sum_{n=0}^{N-1}[2s\ln(1 + z^*_{n + 1} z_n) - 
\\ \nonumber
&&2s\ln(1 + |z_n|^2) - \Delta H(z^*_{n+1},z_n)],
\end{eqnarray}  
where the variables $z_n$ satisfy the periodic boundary 
conditions: 
$ z_N = z_0$ or in the continuum limit $z(T) = z(0)$, and the Hamiltonian
\begin{eqnarray}  \label{5a}
H(z^*_{n+1},z_n) = \frac{\langle z_{n+1}|\hat{H}|z_n\rangle}
{\langle z_{n+1}|z_n\rangle}.
\end{eqnarray}  

 The matrix elements of all essential operators have the 
form
\begin{eqnarray}   \label{6}
&&\langle z|\hat{S}_0|z'\rangle  = -s(1 - z^*z')(1 + z^*z')Q_s, \ \ 
\\ \nonumber
&&\langle z|\hat{S}_-|z'\rangle  = 2sz'(1 + z^*z')Q_s, 
\\ \nonumber  
&&\langle z|\hat{S}_+|z'\rangle  = 2sz^*(1 + z^*z')Q_s, \ \
\\ \nonumber
&&\langle z|\hat{S}^2_+|z'\rangle  = 2s(2s-1)z^{*2}Q_s, 
\\ \nonumber 
&&\langle z|\hat{S}^2_-|z'\rangle  = 2s(2s-1)z'^2Q_s, \ \ 
\\ \nonumber
&&\langle z|\hat{S}^2_0|z'\rangle  = (s^2(1 + z^*z')^2 - 
2s(2s-1)z^*z')Q_s 
\\ \nonumber
&&\langle z|z'\rangle  = (1 + z^*z')^2 Q_s, \ \ 
\\ \nonumber
&&Q_s =(1 + z^*z')^{2s-2}/(1 + |z|^2)^s(1 + |z'|^2)^s.
\end{eqnarray}

\section{Naive classical picture for the ground state energy and its
instanton splitting} \label{sec:Nai}\ \  

\subsection{Classical action and its minima}
In many cases (but not in all !) one can take the 
continuum limit when
$z_{n+1}$ is close to $z_n$. In this continuum limit it 
is convenient to
introduce the canonically conjugated variables $\varphi$ 
and $p = \cos \theta$:
\begin{eqnarray}    \label{7}
z=\rho \exp{(i\varphi)},\ \ \ \rho^2=\frac{s - p}{s+p}, 
\ \ \ \begin{array}{c} 0 \leq \varphi \leq 2\pi, \\  
-s \leq p \leq s \end{array} .
\end{eqnarray}                                
The action of the system $A(z)$ in the continuum limit 
is
\begin{eqnarray}    \label{8}
&&A(\varphi,p) = \int_0^T[i(p - s)\dot\varphi - 
H(p,\varphi)]dt,  \\
&&H(p,\varphi) = (g/2)(p^2 (1+\cos^2\varphi)- 
s^2\cos^2\varphi), 
\end{eqnarray}  
where $H(p,\varphi)$ is the Hamiltonian of the problem, 
$g = f(2S-1)/S, \ \ g > 0$ is the coupling
constant. This Hamiltonian is unusual because the mass 
depends on the coordinate
$\varphi$. The second essential peculiarity is the 
presence in the action of the term 
$is\dot\varphi$ which separates integer and half-integer 
spins \cite{Los,Del}. 
In spite of that the ground state energy and the 
splitting can be found.

  One can easily check that the Hamiltonian has two 
minima at the points $\varphi = 0,\
 \pi$ and $p = 0$.
These minima are deep for $s \gg 1$ and the Hamiltonian 
in the neighborhood of
the minima has the simple form of a harmonic oscillator
\begin{eqnarray}   \label{9}
H(p,\varphi) = E_{min} +gp^2 + gs^2\varphi^2/2, \ \ \ 
E_{min} = -s^2g/2
\end{eqnarray}     
and the vibration frequency is $\omega = \sqrt{2}gs$. 
One can assume that
the energy of the ground state 
\begin{eqnarray}   \label{9a}
E_{0\ naive} = E_{min} + \hbar \omega /2. 
\end{eqnarray} 
This expression for $E_0$ can be obtained if we calculate the
contribution of the trivial saddle piont to the partition 
function by the method of \cite{Col}.

\subsection{The instanton contribution to the splitting of 
the ground state} \ \

 Besides the trivial saddle point contribution to the 
partition function 
discussed above (\ref{5}) there
exists a set of nontrivial saddle point contributions. 
These saddle points, in 
their main features, can be discussed in terms of the 
continuum action (\ref{8}).
The discussion of the Gaussian fluctuations around the 
saddle points has to be
done more accurately but the result can be expressed in 
terms of the continuum
action only. The instanton saddle points for the action 
(\ref{8}) are realized 
at the imaginary momentum $p_t = isu_t$.
The Hamiltonian equations of motion which follow from 
(\ref{8}) are:
\begin{eqnarray} \label{21}
&&\dot{\varphi} = gs(1+\cos^2\varphi)u, \\  \nonumber
&&\dot{u} = gs\sin\varphi \cos\varphi(1 + u^2).
\end{eqnarray} 

 We are interested in the time periodic solution of Eqs. 
(\ref{21}):
$\varphi(0) = \varphi(T), \ p(0) = p(T)$. These periodic 
solutions can be constructed
from the elementary solutions which are named instantons 
and anti--instantons.
It is obvious from Fig. 1 that there are two types of 
instantons (A and B) and 
two types of anti--instantons ($\bar{A}$ and $\bar{B}$) 
for our problem. 
\begin{figure}
\centering
\epsfig{figure=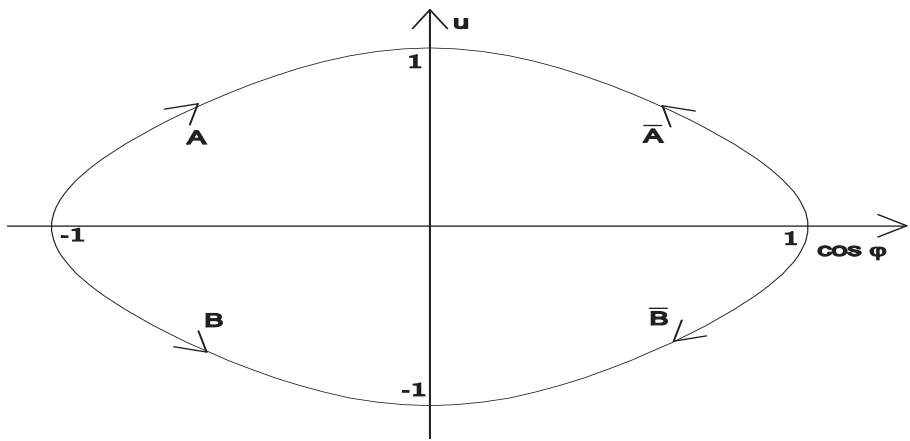,height=6.cm,width=12.cm}
\caption{Instanton trajectories in the phase $cos(\varphi) - u$ space. 
The curve A corresponds to the $A$--instanton and $\bar{A}$ antiinstanton, 
the curve B corresponds to the B-instanton and $\bar{B}$ antiinstanton. 
$cos(\varphi)$ is along the horizontal
axis, the momentum $u$ is along the vertical axis.}
\label{fig1} \end{figure}
Each of them is labeled by the time $t_0$ at
which $\cos(\varphi(t_0)) = 0$ and $u(t_0) = \pm 1$. 

The A--instanton starts at the time $t \rightarrow 
-\infty$ from the point of phase space
$(\varphi = -\pi, u = 0)$ at the time $t = t_0$ arrives 
at the point 
$(\varphi = 0, u =-1)$ and at the time $t \rightarrow 
\infty$ arrives at the 
point $(\varphi = \pi, u = 0)$.  

 The B--instanton starts at the time $t \rightarrow 
-\infty$ from the point of phase space
$(\varphi = -\pi, u = 0)$ at the time $t = t_0$ arrives 
at the point 
$(\varphi = 0, u =1)$ and at the time $t \rightarrow 
\infty$ arrives at the point 
$(\varphi = \pi, u = 0)$.  

The A--anti--instanton starts at the time $t \rightarrow 
-\infty$ from the point of phase space
$(\varphi = \pi, u = 0)$ at the time $t = t_0$ arrives 
at the point
$(\varphi = 0, u =1)$ and at the time $t \rightarrow 
\infty$ arrives at the  point 
$(\varphi = -\pi, u = 0)$.  

 The B--anti--instanton starts at the time $t 
\rightarrow -\infty$ from the point of phase space
$(\varphi = \pi, u = 0)$ at the time $t = t_0$ arrives 
at the  point 
$(\varphi = 0, u =1)$ and at the time $t \rightarrow 
\infty$ arrives at the 
point $(\varphi = -\pi, u = 0)$.  

The  analytic form of these solutions can be easily 
constructed if we take into
account the energy conservation during the process of 
motion 
$H(u, \varphi) =  E_{min}$.
In this case we have the following algebraic connection 
between $p$ and $\varphi$:
\begin{eqnarray}   \label{22}
u^2 = \frac{1-\cos^2\varphi}{1+\cos^2\varphi}.
\end{eqnarray} 
Combining formula (\ref{21}) and (\ref{22}) we get the 
equation of motion for
instantons and anti--instantons:
\begin{eqnarray}   \label{23}
\dot{\varphi} = \pm gs\sqrt{1- \cos^4\varphi},
\end{eqnarray} 
where the sign $\pm$ corresponds to the two types of 
instantons.
This equation of motion can be easily solved and we have 
\begin{eqnarray}   \label{24}
\cos\varphi_t^0 = \pm 
\frac{\sinh(\tau_0)}{\sqrt{2+\sinh^2(\tau_0)}}, \ \ \ 
u_t^0 = \mp \frac{1}{\cosh(\tau_0)},  
\end{eqnarray}                  
where $\tau_0 = \omega(t - t_0)$.

The instanton action $A_0$ counting from the energy 
minimum is
\begin{eqnarray}   \label{25}
&&A_0 = gs^2\int_0^T (1-\cos^2\varphi)dt = 
\\ \nonumber
&&s\int_0^{\pi}\sqrt{(1-\cos^2\varphi)
\over (1+\cos^2\varphi)}d\varphi = 2s\ln(1+\sqrt{2}).
\end{eqnarray} 

 The calculation of the instantons and anti--instantons  
contribution to the 
partition function is based on the idea of an 
approximate saddle point.
The general trajectory for an approximate saddle point 
is shown in Fig. 2.
\begin{figure}
\centering
\epsfig{figure=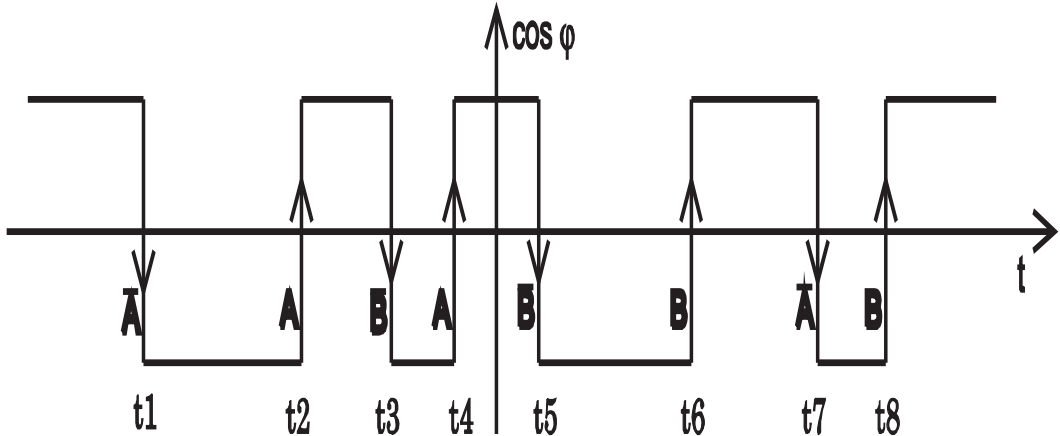,height=6.cm,width=14.cm}
\caption{The scheme of the consecutive fast instanton and
antiinstanton transitions at times $t_1, \ t_2, ... \ t_8$. The time $t$ is
along the horizontal  axis, $\cos(\varphi)$ is along the vertical axis.}
\label{fig2} \end{figure}
This trajectory starts from the point $\varphi = \pi, \ 
u = 0$ and finishes at the 
same one. It represents a sequence of  fast instanton 
and anti--instanton 
transitions (A and B types). The centers of theses 
instantons and
anti--instantons (points where  $\varphi = \pm \pi/2, \ 
u = \pm 1$) are situated
at the arbitrary (but ordered points) $t_l$. Notice, 
that for the exact solution
we have a periodical lattice of instantons and 
anti--instantons. The contribution
of an approximate saddle point to the partition function 
can be presented in
the form (\cite{Col}):
\begin{eqnarray}   \label{26}
&&Z_{apr}^n = \int_0^T \cdot \cdot \cdot \int_0^T 
\prod_{l=1}^{n-1}\theta(t_{l+1} -
t_l)\cdot
\\ \nonumber
&&\prod_{l=1}^{n}[K e^{-A_0 - is\varphi_l} dt_l]
[{\det}'(\hat{L}_{sad})]^{-1/2}.
\end{eqnarray}   
The integral over the positions of the instantons $t_l$ 
in Eq. (\ref{26}) represents
a sum of contributions to the partition function over 
the manifold with almost 
the same action $A_0$. The factor $K^n$ gives
the measure of the integration over degenerate minima 
\cite{Col}. Its explicit
form can be derived if we connect the time $t_l$ with 
the amplitude of the zero mode of the
quadratic form $\hat{L}_{ins}$ describing of the 
fluctuations around of the instanton solution.
The factor  $\exp(-A_0 - is\varphi_l)$ gives  the 
exponent contribution
to the partition function. The quantity $s\varphi_l$  is 
the 
Berry phase which follows from Eq. (\ref{8}) for the 
continuum action. 
This phase $s\varphi_l$ is equal to $s\pi$ for 
$A$--instanton and 
$\bar{B}$--anti--instanton and $-s\pi$ for 
$B$--instanton and $\bar{A}$--anti--instanton.

 The last factor in Eq. (\ref{26}), 
${\det}'(\hat{L}_{sad})$, represents the determinant
of the quadratic form describing fluctuations around the 
approximate saddle
point solution excluding eigenvalues corresponding to 
zero mode
fluctuations. We understand the determinant as a product 
of the eigenvalues of the
corresponding quadratic form. Our aim is to describe the 
calculation of this
determinant in more detail.

 (1). One can represent ${\det}'(\hat{L}_{sad})$ as a 
product
\begin{eqnarray}   \label{27}
{\det}'(\hat{L}_{sad}) = \det(\hat{L}_0)\cdot X,\ \ \     
X = {\det}'(\hat{L}_{sad})/\det(\hat{L}_0),
\end{eqnarray}  
where $\det(\hat{L}_0)$ is the determinant of the 
trivial saddle point which we calculate in Sec. \ref{sec:Det} (\ref{11}). 
It is important to notice that the magnitudes of
determinants  ${\det}'(\hat{L}_{sad})$ and  
$\det(\hat{L}_0)$ depend on the method
of regularization of the functional integral for the 
partition function or
(what is just the same) from the large eigenvalues of 
the quadratic forms
$\hat{L}_{sad}$ and  $\hat{L}_0$. But it is obvious (see 
Fig. 3) that 
for $T\omega/n \gg 1$ the large eigenvalues of the 
quadratic forms
$\hat{L}_{sad}$ and $\hat{L}_0$ are the same because 
during almost all time of motion the variables 
$\varphi(t)$ and $u(t)$ are close
to the trivial saddle points where $\varphi = 0,\ \pi$ 
and $u = 0$. 
\begin{figure}
\centering
\epsfig{figure=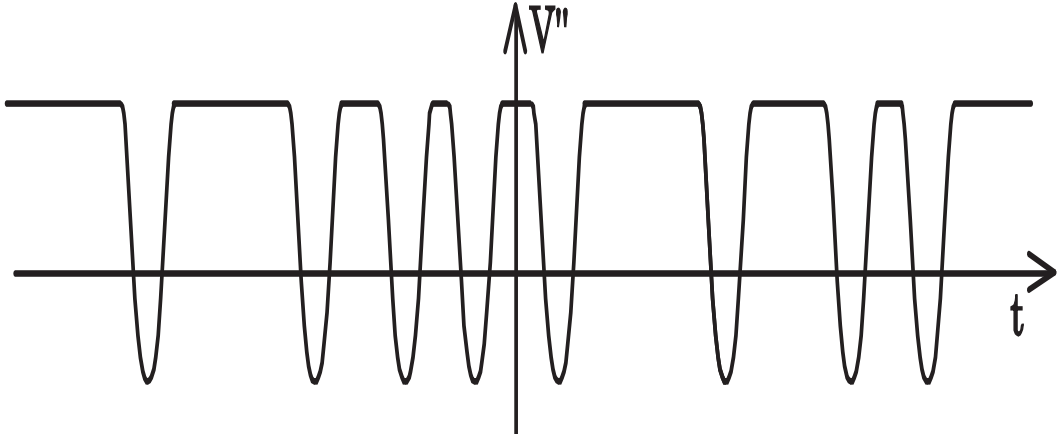,height=6.cm,width=14.cm}
\caption{The effective potential energy for the "particle" in the field of many instanton
transitions. The time $t$ is along the horizontal  axis the effective potential
energy is along the vertical axis.}
\label{fig3} \end{figure}
This means
that the quadratic forms $\hat{L}_{sad}$ and $\hat{L}_0$ 
differ from each
other only during the time $n/\omega \ll T$. This means 
that the quantity $X$ can 
be presented as the ratio of determinants in the 
continuum limit
\begin{eqnarray}   \label{28}
X = {\det}'(\hat{L}_{sad\ cont})/\det(\hat{L}_{0\ 
cont}).
\end{eqnarray}    

 (2). For simplification of the  problem it is 
convenient to integrate the action 
over the variable $p$. For that we spread the limits of 
integration to $\pm \infty$
in the expression (\ref{8}) for the partition function. 
Such spreading of the limits of 
integration can be justified in the case 
\begin{eqnarray}   \label{28a}
g\Delta s^2 \gg 1, \ \ \mbox{or} \ \ s \gg 1/\Delta \omega, 
\ \ \ \Delta \omega \ll 1.
\end{eqnarray} 
Because  $\Delta \omega < 0.1$ this leads to conclusion that
such integration is justified for $s \gg 10$.
We want to stress that this condition includes the time 
quantization step $\Delta$. Therefore the expression for
the partition function takes the form of a functional 
integral over the variable
$\varphi$, and in the leading approximation over $1/s$  we have
\begin{eqnarray}   \label{29}
&&Z = \int_0^{2\pi}\cdot \cdot \cdot\int_0^{2\pi} e^{A_B 
+ A_{\varphi}}
\prod_{t=0}^{T}
\frac{d\varphi_t}{\sqrt{2\pi\Delta 
g(1+\cos^2\varphi_t)}},
\\ \nonumber
&&A_B = -is\int_0^{T}\dot{\varphi}_t\ dt, \ \ \
\\ \nonumber
&&A_{\varphi} = -(1/2)\int_0^{T}\biggl( 
\dot{\varphi}^2_t/
g(1+\cos^2\varphi_t) - gs^2\cos^2\varphi_t\biggr)dt,
\end{eqnarray}    
where $A_B$ is the Berry phase, $A_{\varphi}$ is the 
action of the field $\varphi$. We
want to remind that the measure of integration of the 
functional integral over
configuration space depends on the time step $\Delta$.

 It is natural to introduce a new variable $\psi_t$ in 
the functional integral
(\ref{29}) through relations:
\begin{eqnarray}   \label{30}
&&d\psi_t = (g(1+\cos^2\varphi_t))^{-1/2}d\varphi_t,
\\ \nonumber
&&\psi_t = \int_0^{\varphi_t}(g(1+\cos^2y))^{-1/2}dy.
\end{eqnarray}    
The explicit form of $\psi_t$ as function of $\varphi_t$ 
is an elliptic function
but this explicit form is unnecessary for further 
calculations. In terms of $\psi_t$
the action $A_{\varphi}$ takes the canonical form
\begin{eqnarray}   \label{31}
A_{\varphi} = -(1/2)\int_0^{T}[\dot{\psi}_t^2 - 
gs^2\cos^2\varphi_t]\ dt,
\end{eqnarray}    
where $\varphi_t$ in this formula is a function of 
$\psi_t$. One can check that
this action has the instantons minima for 
$\cos\varphi^0_t$ according to Eq. (\ref{24})
and the instanton action $A_0$ according to (\ref{25}). 
Decomposing the action
(\ref{31}) around the multi-instanton minima and using 
(\ref{30}) we get the part of the action
quadratic in fluctuations  in the form
\begin{eqnarray}   \label{32}
&&A_{\varphi}^q = -(1/2)\int [\delta \psi_t \ 
\hat{L}_{sad\ cont}\ 
\delta \psi_t]\ dt, 
\\ \nonumber
&&\hat{L}_{sad\ cont} = -\frac{\partial^2}{\partial t^2} 
+ V''(\varphi_t^0), \ \ \
\\ \nonumber
&&V''(\varphi_t^0) = \frac{\omega^2}{2} 
(2\cos^2\varphi_t^0 - 1)(1 + \cos^2\varphi_t^0).
\end{eqnarray}    
The function $V''(\varphi_t^0)$ plays the role of the 
potential energy for the problem
of fluctuations around the instantons (see Fig.3). For the trivial 
saddle point
$\cos^2\varphi_t = 1$, $V''(\varphi_t^0)= \omega^2$ and 
the operator 
$\hat{L}_{0\ cont}$   has the simple form:
\begin{eqnarray}   \label{33}
\hat{L}_{0\ cont} = -\frac{\partial^2}{\partial t^2} + 
\omega^2.
\end{eqnarray}   
The typical function $V''(\varphi_t^0)$ is
shown in Fig. 3. It is equal to $\omega^2$ in almost 
all region  
$0 \leq t \leq T$, excluding narrow regions of order 
$1/\omega$ around 
the instanton transitions at points $t = t_l$ where it 
has wells of depth
$-\omega^2/2$.
               
 (3). At this point we can calculate the ratio of 
determinants $X$ (\ref{28}). This
calculation is based on the small parameter $n/\omega T 
\ll 1$ and we can
prove that the ratio $X$ is equal to the product of the 
ratio of determinants
for each region containing an instanton or an 
anti--instanton:
\begin{eqnarray}   \label{34}
X = \prod_{l=1}^{n}X_l, \ \ \ 
X_l = {\det}'(\hat{L}_{l,\ ins\ cont})/\det(\hat{L}_{l,\ 
0\ cont}),                               
\end{eqnarray}     
where the operators  $\hat{L}_{l,\ ins\ cont}$ and 
$\hat{L}_{l,\ 0\ cont}$ are
defined in the region $((t_{l+1} + t_l)/2,\ (t_l + 
t_{l-1})/2)$ around the
position $l$--instanton $t_l$ on the class of functions 
which satisfy the zero
boundary conditions at the points   $((t_{l+1} + t_l)/2$ 
and 
$(t_l + t_{l-1})/2)$.  This statement can be justified 
by the following arguments.
Every determinant $\det(\hat{d})$ is the product of the 
eigenvalues of the operator
$\hat{d}$ on some class of functions. In our case we can 
consider the class of
functions with zero boundary conditions at points 
$((t_{l+1} + t_l)/2$ and 
$(t_l + t_{l-1})/2)$. In this case all eigenvalues will 
be nondegenerated. 
For example, for the periodic boundary conditions they 
will be twice degenerated 
what is inconvenient. All eigenfunctions of the operator 
$\hat{d}$ 
in the region $[0,\ T]$  can be divided into two 
classes: those localized close to
the instanton positions $t_l$ with the "bounded" 
eigenvalues forming almost degenerated groups
of $n$ eigenvalues and the propagating eigenfunctions 
with "continuum" eigenvalues almost the
same as in the case of the trivial saddle point. It is 
obvious that the
contribution of the "bounded" eigenvalues to the total 
determinant can be 
presented as the product of  $n$ contributions. The 
"continuum" eigenvalues can be
calculated in the quasiclassical approximation and can 
also be split into $n$
parts. This argument proves the validity of Eq. 
(\ref{34}).

 (4). Because our problem is reduced to the well-known one 
for the usual instantons we can use results of Coleman
\cite{Col}. The quantities $X_l$ of Eq. (\ref{34}) can be 
calculated if we construct the 
normalized eigenfunction 
$\chi(t) =  N\dot{\psi}(t)$, with $N = 1/\sqrt{A_0}$, of 
the operator 
$\hat{L}_{sad\ cont}$ (\ref{32}) with zero eigenvalue 
\cite{Col}. The function
$\chi(t)$ has a form
\begin{eqnarray}   \label{35}
\chi(t) = -s\sqrt{2g/(A_0(2 + \sinh^2\tau_0))}, \ \ \ \tau_0 =
\omega(t - t_0),
\end{eqnarray}        
with asymptotic behavior at $\tau_0 \Rightarrow \infty$
\begin{eqnarray}   \label{36}
\lim |\chi(t)| \Rightarrow A_{\tau} \exp(-|\tau_0|), \ \ 
A_{\tau} =  2(2)^{1/4}\sqrt{s\omega /A_0}.
\end{eqnarray}   
On the basis of this solution one can prove \cite{Col} 
that the quantity
\begin{eqnarray}   \label{36a}
X_l = 1/(2 A^2_{\tau})
\end{eqnarray}   
and does not depends on $l$ and $t_l$. The final formula 
for ${\det}'(\hat{L}_{sad})$
entering in Eq. (\ref{26}) for the partition function 
$Z_{apr}^n$ can be easily obtained
if we combine Eqs. (\ref{27},\ref{28},\ref{34})
\begin{eqnarray}   \label{37}
{\det}'(\hat{L}_{sad}) =  \det(\hat{L}_{0}) /(2 A^2_{\tau})^n.
\end{eqnarray}

   (5). The quantity $K$ from Eq. (\ref{26}) for the 
partition function
$Z_{apr}^n$, which gives the measure of integration over 
time \cite{Col}, 
can be calculated if we take into account the explicit 
form of the measure
$dG_t$ in the functional integral (\ref{29}) for the 
partition function for $s \gg 1$
\begin{eqnarray}   \label{37a}
dG_t = d\psi_t/\sqrt{2\pi\Delta} \Rightarrow Kdt \ \ 
\mbox{with}\ \ K =\sqrt{A_0/2\pi}.
\end{eqnarray}  

 At present all quantities entering in Eq. (\ref{26}) for 
$Z_{apr}^n$ are defined 
and we have
\begin{eqnarray}   \label{42}
&&Z_{apr}^n = (\det(\hat{L}_0))^{-1/2}(IT)^n\exp(-i\Phi)/n!, \ \ 
\\ \nonumber 
&&I = C \exp(-A_0), 
C = \sqrt{2\Delta}A_{\tau}K = 2\ 2^{1/4}\omega\sqrt{s/\pi},
\end{eqnarray}   
where $\Phi$ is the total Berry phase during all time of 
transitions.
 
For obtaning the final answer for $Z$ it is necessary to 
sum all 
saddle point contributions to the partition function. 
These total saddle point contributions can be presented 
in the form of the following sum:
\begin{eqnarray}   \label{43}
&&Z = Z_0[1 + (A +B)(\bar{A} + \bar{B})T^2/2 +\ ... \  
\\ \nonumber
&&[(A +B)(\bar{A} + \bar{B})T^2]^n/(2n!) + \ ... ].
\end{eqnarray} 
where $Z_0$ is the contribution of the trivial saddle 
point, 
$A, B,\bar{A}, \bar{B}$ are instanton and 
anti--instanton without the determinant
of the trivial saddle point.
The dependence on the Berry phase can be easily 
determined:
\begin{eqnarray}   \label{44}
&&A,\bar{B} = \exp(i\pi s)I, \ \ \  
B,\bar{A} =  \exp(-i\pi s)I, \\ \nonumber
&&(A +B)(\bar{A} + \bar{B}) = 4I^2 \cos^2(\pi s).
\end{eqnarray} 
After these calculations we can sum all terms of the 
expression (\ref{43}) 
in the  form of $\cosh(-T\Delta E/2)$ 
and write the final answer for the energy of the ground 
state and its splitting
by taking into account the explicit form of $Z_0$ 
(\ref{19})
for the trivial saddle point:
\begin{eqnarray}  
\label{45}
&&E_{\pm} =  E_0 \pm  \delta E/2, \ \ \
E_0 =  E_{min} + \omega/2,
\\ \nonumber
&&\delta E = 8\ 2^{1/4}\omega\sqrt{s/\pi}
\exp(-2s\ln(1+\sqrt{2}))\cos(\pi s)],
\end{eqnarray} 
where $E_{min} = -gs^2/2, \ \omega = \sqrt{2}sg, \ g = f(2s-1)/s$.   

We can see that this splitting is exponentially small 
for large $s$ and equals zero
in the case of half--integer spins in full 
agreement with Kramers
theorem \cite{Los,Del}. The cancellation of the 
splitting for the case
of the half-integer spin takes place due to the 
compensation of the contributions of the
instantons of $A$ and $B$ type to the partition function 
$Z$. 

 But, unfortunately, expressions (\ref{45}) for the ground state energy 
$E_0$ and the instanton splitting $\delta E$ are valid only qualitatively.
The correct answer for $E_0$ contains the correction $-3gs/4$
and for the instanton splitting $\delta E$ the additional factor
$1/(1 + \sqrt{2})$. The origin of these corrections is in the subtle nature
of the functional integral which demands accurate treatments when we calculate
functional determinants.

\section{Energy of the ground state or the trivial 
saddle point contribution to the partition function}\label{sec:Gro} \ \  

  The expression $E_0$ for the ground state energy is wrong 
due to an incorrect definition of the product of 
operators at the same point in terms of the variables $\varphi$ 
and $p$.
 The natural variables for the functional integral
(\ref{4}) are the $z$--variables. The trivial saddle 
points in terms of these
variables are $z = \pm 1$. In the neighborhood of the 
saddle points we can
represent the $z$--variables in the form
\begin{eqnarray}   \label{10}
z = \pm 1 + \sqrt{2/s}\ a, \ \ \, z^* = \pm 1 + \sqrt{2/s}\ a^*.
\end{eqnarray}     
The quadratic part of the action close to the saddle 
point is
\begin{eqnarray}   \label{11}
&&A_q(a) = \sum _{n =0}^{N-1} [(a^*_{n+1}a_n - a^*_n 
a_n)-\alpha a^*_{n+1}a_n - 
\\ \nonumber
&&(\beta/2)(a^{*2}_{n+1} + a^2_n)],
\end{eqnarray}     
where $\alpha = 3Tsg/N, \ \beta = Tsg/N$. The 
quadratic form $A_q(a)$ can
be partly diagonalized if we pass to the 
$\omega$-representation for the variables
$a_n, \ a^*_n$:
\begin{eqnarray}   \label{12}
&&a_n = N^{-1/2} \sum _{m = 0}^{N-1} \exp(i\omega_m n)a_m,
\\ \nonumber
&&a^*_n = N^{-1/2} \sum _{m = 0}^{N-1} \exp(-i\omega_m n)a^*_m, 
\end{eqnarray}     
where $\omega_m = 2\pi i m/N$. The action in the 
$\omega$-representation has the 
form:      
\begin{eqnarray}   \label{13}
&&A_q(a) = \sum _{m = 0}^{N-1}[((1 - 
\alpha)\exp(-i\omega_m) -1)a^*_m a_m -
\\ \nonumber
&&\beta (a^*_m a^*_{N-m} + a_m a_{N-m})].
\end{eqnarray}  
The Gaussian integrals with the quadratic form 
(\ref{12}) can be easily performed
and the determinant $D^{-1}$ can be obtained. The 
determinant $D$ is equal to the product 
of the eigenvalues $\lambda_m$ of the quadratic form 
(\ref{12})
\begin{eqnarray}   \label{14}
\lambda_m = \alpha^2 - \beta^2 + 2(1-\alpha)(1 - 
\cos\omega_m ),
\end{eqnarray}  
The partition function $Z$ has the following expression 
in terms of
$\lambda_m$:
\begin{eqnarray}   \label{15}
Z = \exp(E_{min}T) Y,\ \ \ Y = \prod_{m=0}^{N-1} 
\lambda_m^{-1/2}.
\end{eqnarray}                                         
For the calculation of the quantity $Y$ it is convenient 
to represent the eigenvalues  $\lambda_m$ 
in the form 
\begin{eqnarray}   \label{16}
&&\lambda_m = \mu_m \mu_{N-m},\ \ \ 
\mu_m = b + c\exp(i\omega_m), 
\\ \nonumber
&& (b,c) = (1/2)\left(\sqrt{\alpha^2 - \beta^2} \pm 
\sqrt{(2 - \alpha)^2 - \beta^2}\right) 
\\ \nonumber
&&\simeq \pm (1 -\alpha/2 \pm \sqrt{\alpha^2 - \beta^2}/2).
\end{eqnarray}  
In terms of the quantities $\mu_m$ we have for $Y$ 
\begin{eqnarray}   \label{17}
&&Y = \prod_{m=0}^{N-1} \mu^{-1}_m = \exp [-\sum_ {m = 
0}^{N-1}\ln(\mu_m)] =   
\\ \nonumber
&&\exp [-N\ln(b) - \sum_ {m = 0}^{N-1}\ln(1 + 
(c/b)\exp(i\omega_m))].
\end{eqnarray}  
The sum over $m$ in Eq. (\ref{17}) can be easily 
performed if we expand the logarithmic function in a series over $x = 
(c/b)\exp(i\omega_m)$ and take into 
account  only the 
terms $x^p$ with $p = kN$, where $k$ is integer, which 
give nonzero contributions.
The result for $Y$ has the form
\begin{eqnarray}   \label{18}
Y = b^{-N}(1 - (c/b)^N)^{-1}.
\end{eqnarray}     
Using the property $\alpha, \ \beta, \  b, \ c \simeq 1 + 
d/N $ we easily get for
the partition function and the ground state energy
\begin{eqnarray}   \label{19}
&&Z_0 =2\exp[-E_0 T] [1 - \exp(-\omega T)]^{-1}, \ \ \
\\ \nonumber
&&E_0 = E_{min} + (\omega - \omega_0)/2,
\end{eqnarray}     
where $\omega = \sqrt{2}sg, \ \omega_0 = 3sg/2$. The 
factor $2$ in this formula
takes into account the presence of two minima in the 
action or the two types of the
low--frequency excitations.

  We can see that the result (\ref{19}) differs from the 
naive result 
$E_{0\ naive}$ obtained previously. From the point
of view of the functional integral this difference is 
due to the contribution of
the large eigenvalues of the quadratic form (\ref{11}) 
which correspond to
the large $\omega$ in the partition function (\ref{15}). 
From the operator
point of view the quadratic form (\ref{11}) expressed in 
terms of operators (for
example with the help of the Holstein -- Primakoff 
representation with the
quantization along $x$) has to be diagonalized with the 
help of the Bogoluibov's
$u-v$ transformation. As a  result of such diagonalization 
the ground state energy
is shifted by the amount  $(\omega - \omega_0)/2$. 

\section{Calculation of the determinant  and the measure of integration over
time in the continuum approximation}\label{sec:Mea} 

\subsection{Calculation of the determinant} \ \

In this section we present the calculation of the ratio of the functional 
determinants (\ref{24}) $X_l$ without the primary integration over the variable
$p$ which permits us to avoid the artificial limitation (\ref{28a})
$g\Delta s^2 \gg 1$. Our consideration is based on the 
arguments of Coleman \cite{Col} applied to the system of two
differential equations of the first order in time. This system of equations
can be easily obtained if we introduce the quantum corrections 
to the classical solution
$\varphi_t^0, \ u_t^0$ (\ref{24})
\begin{eqnarray}   \label{A1}
\psi_t = \sqrt{s}\delta \varphi_t,\ \ \ \
v_t = \sqrt{s}\delta u_t.
\end{eqnarray}
The quantum correction to the action has the form
\begin{eqnarray}   \label{A2}
A_2 = -(1/2)\int_{-T/2}^{T/2} y_{t i}\hat{L}^{ij}_{0 \ cont} y_{t j} dt,\ \ \ i,j = 1,2,
\end{eqnarray}  
where $y_{t 1} = \psi_t, \ y_{t 2} = v_t$ and the $2 \times 2$ matrix 
$\hat{L}_{0}$ can be presented in the form
\begin{eqnarray}   \label{A3}
&&\hat{L}_{0} = -
\hat{\epsilon}\frac{\partial}{\partial t} + \hat{B}, \ \ \
\hat{\epsilon}^2 = -1,
\\ \nonumber
&&\hat{\epsilon} = \left( \begin{array}{cc} 0 & 1 \\ -1 & 0 \end{array} \right), 
\ \ \ 
\hat{B} =  \left( \begin{array}{cc} a_t & b_t \\ b_t & c_t \end{array} \right), 
\end{eqnarray}  
where the functions $a_t,\ b_t,\ c_t$ can be easily calculated if we take
the variation of Eq. (\ref{21}) with respect to $\varphi_t$ and $u_t$: 
\begin{eqnarray}   \label{A3a}
&&a_t = gs(1+u_t^2)\cos(2\varphi_t), \ \ \ b_t = gsu_t \sin(2\varphi_t),
\\ \nonumber
&&c_t = -gs(1 + \cos^2(\varphi_t)). 
\end{eqnarray} 
For the trivial saddle point we have
\begin{eqnarray}   \label{A3b}
a_t = gs, \ \ \ b_t = 0,\ \ \ c_t = -2gs. 
\end{eqnarray} 
One can see that the operator $\hat{L}_{0}$ is Hermitian.  

  We can fix the class of eigenfunctions
of the operator $\hat{L}_{0}$  (\ref{A3}) if we demand that 
$\psi(-T/2) = \psi(T/2) =0$. For this class of eigenfunctions all eigenvalues
of the operator $\hat{L}_{0}$ are nondegenerated and the ratio of the
determinants $X_l$ for the instanton saddle point and for the trivial saddle
point is the ratio of the products of all eigenvalues. The same arguments as in
\cite{Col} leads us to the conclusion that
\begin{eqnarray}   \label{A4}
X_l =\psi^0_{inst}(T/2)/\psi^0_{triv}(T/2) \lambda_0.
\end{eqnarray}       
Here $\psi^0_{inst}(t)$ and $\psi^0_{triv}(t)$ are eigenfunctions of the
operator $\hat{L}_{0}$ (with zero eigenvalue) with the boundary conditions:
\begin{eqnarray}   \label{A4a}
\psi^0(-T/2) = 0 \ \mbox{and} \ \partial\psi^0(-T/2)/\partial t = 1,
\end{eqnarray} 
and $\lambda_0$ is the eigenvalue of the operator $\hat{L}_{0}$ which
corresponds to zero mode.

  For the case of
the trivial saddle point such solution can be constructed directly:
\begin{eqnarray}   \label{A5}
&&{\bf y}(t) \equiv (\psi(t), v(t)) = 
\\ \nonumber
&&\omega^{-1} (\sinh(\omega (t + T/2)),
\cosh(\omega (t + T/2)/\sqrt{2}),
\end{eqnarray}  
where $\omega = \sqrt{2}sg$ and 
\begin{eqnarray}   \label{A6}
\psi^0_{triv}(T/2) \simeq \exp(\omega T)/2\omega.
\end{eqnarray}

  For the case of the instanton saddle point the function $\psi^0_{inst}(t)$
can be constructed on the basis of the zero mode eigenfuction of the operator
$\hat{L}_{0}$ which is proportional to the time derivative of the instanton
solution (\ref{24})
\begin{eqnarray}   \label{A7}
&&{\bf y}_t^a = \sqrt{s} A (\dot{\varphi}_t^0,\ \dot{u}_t^0) =
\\ \nonumber
&&2\sqrt{2} \omega A\left(-\frac{2\cosh(\tau_0)}{1+\cosh^2(\tau_0)}, \
 \frac{\sqrt{2}\sinh(\tau_0)}{\cosh^2(\tau_0)}\right),
\end{eqnarray} 
where $A$ is the normalization constant due to the normalization condition
\begin{eqnarray}   \label{A8}
\int^{\infty}_{-\infty}  {\bf y}_t^2 dt = 1.
\end{eqnarray} 
This solution has the simple asymptotic form
\begin{eqnarray}   \label{A9} 
{\bf y}_t^a = 2\omega A(-\sqrt{2}, \ \pm 1)\exp(-|\tau_0|) \ \ \ 
\mbox{for} \ \ \ |\tau_0| \Rightarrow \pm \infty.
\end{eqnarray} 
The solution (\ref{A7}) does not satisfy the boundary conditions (\ref{A4a})
and we have to construct the second linear independent eigenfunction 
${\bf y}_t^b$ of the operator  $\hat{L}_{0}$  with the zero eigenvalue.
The asymptotic form of this solution can be easily determined from the solution
for the trivial saddle point
\begin{eqnarray}   \label{A10} 
{\bf y}_t^b = 2\omega A(\mp\sqrt{2}, \ -1)\exp(|\tau_0|) \ \ \ 
\mbox{for} \ \ \ |\tau_0| \Rightarrow \pm \infty.
\end{eqnarray} 

 It is important to note that these two solutions satisfy at any $t$ the condition of
conservation of the current or the Wronskian which can be easily deduced from the
explicit form of the operator $\hat{L}_{0}$
\begin{eqnarray}   \label{A11} 
I = {\bf y}_t^a \hat{\epsilon} {\bf y}_t^b = 16\omega A^2.
\end{eqnarray} 

 At this point we can construct the eigenfunction $\psi^0_{inst}(t)$
\begin{eqnarray}   \label{A12} 
{\bf y}^0_{inst}(t) = -(4\sqrt{2} A \omega^2)^{-1}(e^{\omega T/2}
{\bf y}_t^a + e^{-\omega T/2}{\bf y}_t^b).  
\end{eqnarray}  
One can easily check that $\psi^0_{inst}(T/2) = 1$.

 The last step which is necessary in order to determine the ratio $X_l$
consists of finding the eigenvalue $\lambda_0$ which corresponds to the zero
mode. For that it is convenient to convert the equation on the eigenvalues
$\lambda$ into the form of an integral equation. This integral equation has the 
form
\begin{eqnarray}   \label{A14} 
{\bf y}^{\lambda}(t) = &&{\bf y}^0_{inst}(t) + \frac{\lambda}{16 \omega^2A^2}
\int_{-T/2}^t [{\bf y}_t^a(({\bf y}^b_{t'}\cdot {\bf y}^{\lambda}(t')) - 
\\ \nonumber
&&{\bf y}_t^b ({\bf y}^a_{t'} \cdot {\bf y}^{\lambda}(t'))] dt'.
\end{eqnarray}  
Using conservation of the current $I$ (\ref{A11}) one can verify that this
equation is equivalent to the original one. Moreover, the boundary condition
$\psi^{\lambda}(-T/2) = 0$ is satisfied. The second boundary condition
in the framework of perturbation theory has the form
\begin{eqnarray}   \label{A15} 
0 \equiv \psi^{\lambda}(T/2)=&& 1 + \frac{\lambda}{16 \omega^2A^2}
\int_{-T/2}^{T/2} [\psi_t^a(({\bf y}^b_{t'}\cdot {\bf y}^0_{inst}(t')) - 
\\ \nonumber
&&\psi_t^b ({\bf y}^a_{t'} \cdot {\bf y}^0_{inst}(t'))] dt'.
\end{eqnarray}  
The integral in the right hand side of Eq.(\ref{A15}) can be calculated
asymptotically \cite{Col} and we obtain the following relation for determination of
$\lambda_0$
\begin{eqnarray}   \label{A16} 
0 = 1 - (\lambda_0/16\sqrt{2} \omega^2 A^2 \omega)\exp(\omega T).
\end{eqnarray}  
Combining Eqs. (\ref{A4},\ref{A6},\ref{A16}) we get for the ratio $X_l$ (\ref{A4})
\begin{eqnarray}   \label{A17} 
X_l = &&\frac{1}{(\exp(\omega T)/2\omega)\cdot(\Delta 16\sqrt{2} \omega^3
A^2)\exp(-\omega T)} = 
\\ \nonumber
&&\frac{1}{8\sqrt{2}\omega^2\Delta A^2}.
\end{eqnarray}  
We take into account in Eq. (\ref{A17}) that the operator in the quadratic form
(\ref{A2}) is in fact proportional to $\Delta$ and all eigenvalues are also
proportional to $\Delta$.

\subsection{Calculation of the measure of integration over time} \ \

 The measure of integration over time can be determined if we follow
basically \cite{Col}. The difference between our approach and \cite{Col}
is that we use the canonical formulation. The measure of integration over
variables $\psi$ and $v$ in the leading order (with respect to $1/s$)
is canonical
\begin{eqnarray}   \label{A18} 
d\mu = \prod_{n=0}^{N-1} \frac{d\psi_n}{\sqrt{2\pi}}\frac{dv_n}{\sqrt{2\pi}}, \ \ \ 
\Delta N = T.
\end{eqnarray}  
We stress in this expression for the measure that integration over each
independent variable contains a factor $1/\sqrt{2\pi}$. We want to pass to
integration over the amplitude of the eigenfunctions of the operator
$\hat{L}_{0}$ (\ref{A3}). Then the measure takes the form
\begin{eqnarray}   \label{A19} 
d\mu = \prod_{i=0}^{2N-1} \frac{dc_i}{\sqrt{2\pi}}.
\end{eqnarray}  
We want to find the connection between the amplitude of the zero mode $c_0$ and the
time variable $t_0$ which corresponds to the shift of position of the instanton.
For that let us calculate the change of the original variables 
${\bf y}_t = (\psi_t, \ v_t)$ due to the change of $c_0$ and $t_0$:
\begin{eqnarray}   \label{A20} 
&&d{\bf y}_t = \sqrt{\Delta}{\bf y}_t^a dc_0, 
\\ \nonumber
&&d{\bf y}_t = \sqrt{s} (\dot{\varphi}_t^0,\ \dot{u}_t^0) dt =
A^{-1} {\bf y}_t^a dt.
\end{eqnarray}  
The additional factor $\sqrt{\Delta}$ preceding $dc_0$
is due to the difference between continuos and discrete normalization.
Comparing these two expressions we get

\begin{eqnarray}   \label{A21} 
d\mu_0 = dc_0/\sqrt{2\pi} = K dt,\ \ \ 
K =  A^{-1} \sqrt{s/2\pi\Delta}.
\end{eqnarray}  
 Therefore the constant $C$(\ref{42}) which determines the instanton contribution
to the partition function is
\begin{eqnarray}   \label{A22} 
C \equiv  X_l^{-1/2}K =  2 2^{1/4}\sqrt{s/\pi} \omega.
\end{eqnarray} 
The normalization constant $A$ completely disappears from the final expression for
the constant $C$.
This expression completely coincides with that one (\ref{42}) which was
obtained on the basis of paper \cite{Col} for the action integrated out over
the variable $p$.

\section{Calculation of corrections to the determinant caused by the discrete
nature of the functional integral over time}\label{sec:Det} \ \

 In our calculation we passed carelessly from the original variables 
$z^*_t,\ z_t$ to the natural variables $\varphi_t, \ p_t$. Small corrections
of the order $\Delta \omega$ can lead to finite contributions to such
quantities as the measure of integration and $\det(\hat{L}_{0})$. Practically
they are the same quantities because by change of variables 
of integration they can be
reduced to each other. At first we explain a mechanism of such corrections on a
qualitative level and after that we apply it to our case.

 Let us suppose that each variable $c_i$ in the measure (\ref{A19}) has a small
multiplicative correction of the order $\Delta \omega$
\begin{eqnarray}   \label{A23} 
&&d\mu \Rightarrow \prod_{i=1}^{N} \frac{dc_i}{\sqrt{2\pi}}
(1 + \alpha_i \Delta \omega) = d\mu_0 R, \ \ \ 
\\ \nonumber
&&R = \prod_{i=1}^{N}(1 + \alpha_i \Delta \omega). 
\end{eqnarray} 
This factor $R$ to $d\mu_0$ can be presented in an exponential form and it
is different from unity
\begin{eqnarray}   \label{A24} 
&&R =\exp(\delta A_0),\ \ \ \
\\ \nonumber
&&delta A_0 = \sum_{i=1}^{N}\ln(1 + \alpha_i \Delta \omega) \simeq
\omega\int_0^T \alpha_t dt, 
\end{eqnarray} 
where $\delta A_0$ is the correction to the classical action.
If the function $\alpha_t \rightarrow const$ at $t \rightarrow T$ then
$\delta A_0 \sim const \ \omega T$. If the function $\alpha$ is localized over time
within a region order $\omega^{-1}$ then $\delta A_0 \sim const\ $. 

 If we calculate the determinant the mechanism of influence of
corrections is more subtle. Let us suppose that the matrix  $\hat{L}$ can be
presented in a form
\begin{eqnarray}   \label{A25} 
\hat{L} = \hat{L}^0 + \Delta \omega \hat{Q}, 
\end{eqnarray} 
where the operator $\hat{L}^0$ represents some discrete version
of the operator $\hat{L}_{0}$ of (\ref{A3}) and the operator $\hat{Q}$
represents some discrete version of the second order differential operator
\begin{eqnarray}   \label{A26} 
\hat{Q} = a_0 + a_1\omega^{-1}\frac{\partial}{\partial t} + 
a_2\omega^{-2}\left(\frac{\partial}{\partial t}\right)^2,
\end{eqnarray} 
where functions $a_0,\ a_1, \ a_2 $ are some smooth functions of time $t$
of the order of unity.
Simple arguments shows that the contribution to the determinant due to the 
$a_0$ term
is always small contrary to the contributions of the $a_1$ and $a_2$ terms which 
are of order of unity. To demonstrate this statement let us represent
the operator $\hat{L}$ in the form 
\begin{eqnarray}   \label{A27} 
\hat{L} = [1 + \Delta \omega \hat{Q} \hat{G}^0]\hat{L}^0, \ \ \
\hat{G}^0 = (\hat{L}^0)^{-1},
\end{eqnarray} 
and the determinant takes a form
\begin{eqnarray}   \label{A28} 
&&\det[\hat{L}] \simeq \det[\hat{L}^0]\ \exp(-2\delta A_0), \ \ \
\\ \nonumber
&&\delta A_0 = -(1/2) \Delta \omega\mbox{Tr} [\hat{Q}\hat{G}^0].
\end{eqnarray} 

  One can check that in the discrete representation the matrix elements of the Green's 
function $(\hat{G}^0)_{nn'}$ are of the order $\Delta$ and 
the characteristic width over
$n-n'$ is of order $(\omega \Delta )^{-1}$. This means that the contribution of the $a_0$
terms of the operator $\hat{Q}$ into $\delta A_0$ is of order $\Delta$. This
can easily be checked if we convert the trace in Eq. (\ref{A28}) into an integral
over $t$.

 The first impression is that the contributions of the $a_1$ and $a_2$ terms
to $\det[\hat{L}]$ are also small because the Green's function  $(\hat{G}^0)_{nn'}$ 
has a width over $t-t'$ of order $\omega^{-1}$ and its derivatives have the order
of magnitudes $\omega$. But these arguments are completely wrong because the
Green's function is a singular function over $t-t'$. It has a jump at $t = t'$
and the time derivative of this jump is of order $\Delta^{-1}$. 
  If the functions $a_1$ and $a_2$ $\rightarrow const$ at $t \rightarrow T$ then
$\delta A_0 \sim const \ \omega T$. If the functions $a_1$ and $a_2$ 
are localized over time in a region of order $\omega^{-1}$ then $\delta A_0 \sim const\ $. 

 At this point we can calculate corrections to the instanton contribution due to
renormalization of the measure and the functional determinant in the leading
order with respect to $1/s$. There are three sources of corrections of order
$\Delta$ to the continuos approximation. The first origin of corrections is the
Berry phase. The action with sufficient accuracy is of the form
\begin{eqnarray}   \label{A29} 
A_B = s\sum_{n=0}^{N-1}\left[2\frac{(z^*_{n+1}- z^*_n)z_n}{1+z^*_nz_n} -
\frac{(z^*_{n+1}- z^*_n)^2z_n^2}{(1+z^*_nz_n)^2}\right].
\end{eqnarray} 
The second origin of corrections is the Hamiltonian due to its dependence  on
$z^*_{n+1}$:
\begin{eqnarray}   \label{A30} 
A_{H1} =gs^2\Delta\sum_{n=0}^{N-1}
\frac{(z^*_{n+1}- z^*_n)(3z_n + z^*_n - 3z^*_nz_n^2 - z_n^3)}{(1+z^*_nz_n)^3}.
\end{eqnarray} 
The third origin of corrections is the difference between the average of the
square of the Hamiltonian and the square of the of the average of the  
Hamiltonian:
\begin{eqnarray}   \label{A31} 
A_{H2} =(1/2)\Delta^2\sum_{n=0}^{N-1}\left[
\frac{\langle z_n|\hat{H}^2|z_n\rangle} 
{\langle z_n|z_n\rangle} -
\left(\frac{\langle z_n|\hat{H}|z_n\rangle}  
{\langle z_n|z_n\rangle}\right)^2\right]     
\end{eqnarray} 
Because this correction depends (in the leading order) only on $z^*_n$ and
$z_n$ it can contribute only to the renormalization of the measure of integration.

 At this stage we can produce all calculations of the measure and the
determinant for the instanton contribution with the necessary accuracy. For
that we begin by interpreting the integration over the variables $z'_n$ and $z''_n$
($z_n = z'_n + i z''_n$ and  $z^*_n = z'_n - i z''_n$) 
in expression (\ref{4}) for the partition function $Z$ as an integral over
the two-dimensional surface $\mbox{Im}(z'_n) = 0$ and $\mbox{Im}(z''_n) = 0$ of the 
four dimensional space where variables $z'_n$ and $z''_n$ are considered as 
complex variables. Because the function $(1+z^*_nz_n)^{-2}\exp(A(z^*_n,z_n))$ is
an analytic function of the variables $z'_n$ and $z''_n$ it can be continued in the
four-dimensional complex manifold and in this way we can arrive to the instanton saddle 
point. In the neighborhood of the saddle point we have to integrate over the 
two-dimensional manifold which realizes the directions of steep descent. 
The direction of steep descent is chosen correctly if all eigenvalues of
the quadratic form in the exponent of the action are real.

 This program can be realized with the help of the following change of
variables $z^*_n$ and $z_n$ in the neighborhood of the saddle point:
\begin{eqnarray}   \label{A32} 
&&z_n = \bar{z}_n + (i/\sqrt{s})\bar{z}_n(\psi_n - v_n/(1+u^2_n)),
\\ \nonumber
&&z^*_n = \bar{z}^*_n - (i/\sqrt{s})\bar{z}^*_n(\psi_n + v_n/(1+u^2_n)).
\end{eqnarray}
Here we understand $\bar{z}_n$ and $\bar{z}^*_n$  as classical (nonfluctuating)
variables connected with the previously introduced variables $\varphi_n$ and
$u_n$ by the relations:
\begin{eqnarray}   \label{A33} 
&&\bar{z}_n(\varphi_n,u_n) = \sqrt{(1-iu_n)/(1+iu_n)}\exp(i\varphi_n),
\\ \nonumber
&&\bar{z}^*_n(\varphi_n,u_n) = \sqrt{(1-iu_n)/(1+iu_n)}\exp(-i\varphi_n),
\end{eqnarray}
thus the variables  $\bar{z}_n$ and  $\bar{z}^*_n$ are not in our case complex
conjugated. We shall understand that the classical variables  $\bar{z}_n$ and 
$\bar{z}^*_n$ 
or ${\bf x}_n =(\varphi_n, \ u_n)$ satisfy the classical equations of motion which
determine the saddle point with  corrections of order $\Delta$ taken into account. 
This means that only in the main approximation the variables ${\bf x}$
are equal to  ${\bf x}^0_n = (\varphi^0_n, \ u^0_n)$ determined by Eq. (\ref{24}). The
difference between variables ${\bf x}$ and variables ${\bf x}^0$ is of order 
$\Delta$ and is determined by small corrections of order of $\Delta$
contained in the action $A_B$ (\ref{A29}) and corrections to the action
$A_{H1}$ (\ref{A30}) and $A_{H2}$ (\ref{A31}). It is fortunate that these
corrections are nonessential to our problem. 

  The reason for that is the canonical form of the measure (\ref{A18}) in terms
of the variables ${\bf y}_n = (\psi_n, \ v_n)$. Notice that the change of variables
(\ref{A32}) was stimulated by formulas of differentiation over
time of the quantities $\bar{z}_n(\varphi_n,u_n)$ and $\bar{z}^*_n(\varphi_n,u_n)$.

 At this stage we can calculate the renormalization of the functional determinant
due to the $\Delta$-corrections. This can be done on the basis of the following
formula for the decomposition of the action in the neighborhood of the
saddle point:
\begin{eqnarray}   \label{A35}
A(z^*,z) = A_0 + \frac{1}{2s}\sum_{ni,n'j}\frac{\partial^2 A(\varphi, u)}
{\partial x_{ni} \partial x_{n'j}}\ y_{ni} y_{n'j} + ...\ .
\end{eqnarray}
This formula strongly simplifies the calculations and can be proved if we 
consider the original variables $z_n$ and $z^*_n$ as function of variables 
${\bf y}_n = (\psi_n, \ v_n)$ from one hand and the variables  
${\bf x}_n =(\varphi_n, \ u_n)$ from another.

  The matrix $\hat{L}^0$ can be chosen on the basis  (\ref{A35}) in the form
\begin{eqnarray}   \label{A36}
\hat{L}^0 = \left( \begin{array}{cc} a_n,& -\partial_- +b_n \\
\partial_+ +b_n, & c_n \end{array} \right),
\end{eqnarray}
where the functions $a_n, \ b_n$ and $c_n$ were determined in Eq. (\ref{A3a}),
and the difference derivatives are determined by the following relations:
\begin{eqnarray}   \label{A37}
&&\partial_-f_n = (f_n - f_{n-1})/\Delta, \ \ \
\partial_+f_n = (f_{n+1} - f_n)/\Delta, \ \ \ 
\\ \nonumber
&&\partial^2 f_n = (f_{n+1} + f_{n-1} -2 f_n)/\Delta^2.
\end{eqnarray}
The Green's function $\hat{G}^0$ 
satisfies the relation:
\begin{eqnarray}   \label{A38}
\sum_{n_a}(\hat{L}^0)_{n,n_a}(\hat{G}^0)_{n_a,n'} = \delta_{n,n'},
\end{eqnarray}
and can not be found in general form. But this is unnecessary for our purposes.
We are interested in  the singular part of the Green's function $(\hat{G}^0)_{n'n}$ at 
$n' \approx n$. This singular part of the Green's function at $n'\approx n$ can
be found in a general form:  
\widetext
\begin{eqnarray}   \label{A39}
(\hat{G}^0)_{n',n} = \Delta  \left( \begin{array}{cc} 
\Delta \theta_{n',n}(n'-n)c_n,& \theta_{n',n}(1 - \Delta (n'-1 -n)b_n) \\
-\theta_{n'+1,n}(1 + \Delta (n'+1-n)b_n), & \Delta \theta_{n'+1,n}(n'-n)a_n 
\end{array} \right),
\end{eqnarray} 
\narrowtext
\noindent
here $\theta_{n'n}$ is the $\theta$-function defined in the following manner
\begin{eqnarray}   \label{A40}
\theta_{n'n} =  \left\{ \begin{array}{cc} 1   , & n'\geq n+1 \\
0   , & n' \leq n  \end{array} \right. ,  \
\begin{array}{c} \partial_+\theta_{n',n} =\Delta^{-1}\delta_{n',n}\\

\ \partial_-\theta_{n'+1,n} =\Delta^{-1}\delta_{n',n} \end{array}.
\end{eqnarray}

 After some tedious calculations the singular (containing essential derivatives) part of
the operator $\hat{Q}$ entering in Eq. (\ref{A27}) can be presented in a form:
\widetext
\begin{eqnarray}   \label{A41}
(\hat{Q}f)_n = (1/2 \omega)\left( \begin{array}{cc} -(1+u_n^2)\partial^2  , & 
\partial^2 + (2 \dot{\varphi}_nu_n - 3gs)\partial_-\\
\partial^2 -(2 \dot{\varphi}_nu_n - 3gs)\partial_+ , &(1+u_n^2)^{-1}\partial^2
\end{array} \right) f_n , \ \ \ \
\end{eqnarray}
\narrowtext
\noindent
where $f_n$ is an arbitrary function.  For the diagonal elements of the matrix
$\hat{Q}$ it is sufficient to keep the second derivative only. For nondiagonal
elements we have to keep the first and the second derivatives. 
Acting with the operator $\hat{Q}$ on the
Green's function $\hat{G}^0)_{n',n}$ and calculating the trace we get for the
correction to the action $\delta A_0$
\begin{eqnarray}   \label{A42}
&&\delta A_0 = -(1/2)\int (f_t - f_{\infty}) dt, \ \ \ 
\\ \nonumber
&&f_t = (1+u_t^2)^{-1}a_t -(1+u_t^2)c_t + 4\dot{\varphi}_tu_t.
\end{eqnarray}
Since we are interested in the ratio of determinants we subtract from the 
function $f_t$ its value at the trivial saddle-point $f_{\infty}$. 
Using Eq. (\ref{23}) for $\dot{\varphi}$ we get the final expression 
for the correction to the instanton action $\delta A_0$
\begin{eqnarray}   \label{A43}
\delta A_0 = -gs\int_{-\infty}^{\infty} (1- \cos^2(\varphi_t))dt = -\ln(1+\sqrt2).
\end{eqnarray}
 The obtained result is surprisingly simple. It can be found if we change in the
expression for the energy splitting (\ref{45}) $s  \Rightarrow s + 1/2$.
Such change has a quasiclassical meaning and can be found by 
a simpler manner then in this section. Finaly,l result we have instead of the
expression (\ref{45}) for the ground state energy $E_0$ and for the instanton
splitting
\begin{eqnarray}  \label{42a}
&&E_0 =  -s(s+1-\sqrt{2})f,\ \
\\ \nonumber
&&\delta E = 16\ 2^{3/4}\pi^{-1/2}s^{3/2}f
(1+\sqrt{2})^{-(2s+1)}\ \cos(\pi s).
\end{eqnarray} 
This result completely coinsides with the result of \cite{Enz} for the case
$A=B=f$.

\section{Comparison with the numerical results}\label{sec:Num} \ \

In this section we will compare the exact ground-state 
energy and splitting
with the results obtained in the above discussion, 
namely eq(\ref{45}). The Hamiltonian (\ref{1}) is easily 
diagonalized on 
the basis of the eigenfunctions of $\hat S^2$ and $\hat 
S_0$. 
The ground--state belongs to 
the symmetric representation and, so, we will just 
consider this 
representation for different values of the total spin 
$s$. 
In table I the numerical results and the theoretical 
ones
for the ground-state energy (\ref{19}) and its splitting
(\ref{45}) are presented. As expected the relative error 
between the 
calculated and the exact ground-state energy decreases 
with $s$.
 
\section{Discussion}\ \

We want to discuss two points here: the interpretation of the effective change of
$s \rightarrow s + 1/2$ in the effective action and the procedure of
calculation of corrections to the instanton approximation.

\subsection{Interpretation of renormalization of the instanton action}\ \

 Let consider the explicit form of the spin operators acting on the space of
functions 
\begin{eqnarray}   \label{C1}
\psi_m(\varphi) = e^{im\varphi}/\sqrt{2\pi}, \ 
0 \le \varphi \le 2\pi, \  m = -s,...,s.
\end{eqnarray}
The spin operators have the form \cite{Enz,Kle}
\begin{eqnarray}   \label{C2}
&&\hat{S}_+=\sqrt{(s+1/2)^2-(\hat{p}-1/2)^2}\ e^{i\varphi}, \ \ \
\hat{S}_z =\hat{p}, 
\\ \nonumber
&&\hat{S}_-= e^{-i\varphi} \sqrt{(s+1/2)^2-(\hat{p}-1/2)^2}, \ \ \
\hat{p} = -i\partial/\partial \varphi.
\end{eqnarray}
Substituting this representation for the spin operators (\ref{C2})
into the Hamiltonian (\ref{1}) and decomposing it over $\hat{p}$
(this decomposition can be justified) we get in the leading 
approximation over $1/s$:
\begin{eqnarray}   \label{C3}
\hat{H}(\hat{p},\varphi) = (f/2)(\hat{p}^2 (1+\cos^2\varphi)- 
(s + 1/2)^2\cos^2\varphi).
\end{eqnarray}
We can calculate with this Hamiltonian the partition function applying the 
procedure of the $p-\varphi$ construction of the functional integral.
In result we get the "functional
integral" in which in each time section we have an integration over $\varphi$
in the limits $[0,\ 2\pi]$ and a summation over $m$ in the limits $[-s, \ s]$. We can 
change summation over $m$ on integration over $p$ and prolong it to infinty.
After integration over $p$ we get the action which coincides with the action
(\ref{29}) with one essential difference that instead of $s^2$ in (\ref{29})
we have $(s+1/2)^2$ in (\ref{C3}). This means that we can get the correct
answer for the tunneling splitting $\delta E$ (\ref{42a}) in the continuous 
representation \cite{Enz}. Because we know at present that corrections are small (see
discussion below) there is one question: why are corrections absent in the approach
discussed in this section? The explanation lies in the difference between the
coherent state construction of the functional integral and the $p-\varphi$
construction. One can check that at the calculation of the functional
determinant in the $p-\varphi$ functional integral the corrections of the
order $\Delta$ are absent. These corrections are also absent in the usual
problem of tunneling in quantum mechnics which is confirmed by the coincidence
of the result of the energy splitting with the quasiclassical one \cite{Col}).

\subsection{Corrections to the leading approximation}\ \

We can construct perturbation expansion around of the trivial saddle point
(\ref{10}) as well as around of the instanton solution (\ref{A32}). Both these
expansions are over very well defined small parameter $1/s$. This parameter
appears due to the presence of the chracteristic factor 
$1/\sqrt{s}$ in the change of
the variables (\ref{10}) and (\ref{A32}). We want to stress that formulation in
terms of the functional for the spin coherent states gives the regular procedure
for the calculation of corrections. These corrections are of order $const$ for
the calculation the ground state enrgy and are of order $1/s$ for calculation of
the factor before the exponent in the expression for the instanton splitting.

 In conclusion we want to stress one peculiarity of this perturbation theory:
the presence in it of characteristic terms proportional of the number of the time
sections $N$. The presence of such terms is the characteristic feature of the
perturbation theory when the measure of the integration is not trivial. In such
theories the kinetic term is also nontrivial: the effective
mass or the coefficient before a term  with the time derivative  is a function
of the field variables. At the calculation in the framework of perturbation
theory such terms lead to divergencies at large frequences $\omega$. These
divergencies have to compleatly canceal $N$-terms which follow from the
measure.
  
\vskip 1.cm

\acknowledgments

We are grateful to K.A. Kikoin for valuable suggestions and to A. Viera.
This work was supported in part by the 
Portuguese projects N PRAXIS/2/2.1/FIS/451/94,
V. B. was supported in part by the Portuguese program PRAXIS XXI /BCC/ 4381 /
94, and in part by the  Russian Foundation for 
Fundamental Researches, Grant No 94-02-03235.

\vskip .5cm
\mediumtext
\begin{table}
\caption{Results for the ground state energy and its 
splitting.
In the first column is the magnitude of spin, the second 
and third columns 
are, respectively, the exact and
the calculated ground-state energy, the fourth and the 
fifth columns give the exact and the calculated ground-state 
splitting.\label{tab1}}  \vskip 1.cm  \noindent
\begin{tabular}{|c|c c c|c c c|}
\hline\rule[-2mm]{0mm}{7mm}
  $ s $ & $ \quad E/f \quad$ & $\quad E_0/f \quad$ & $\quad E_0/E\quad$ & $\quad
 \Delta E/f \quad$ & $\quad E_{inst}
/f \quad$ & $\quad E_{inst}/\Delta E \quad$ \\ 
  \hline\rule[-2mm]{0mm}{7mm}
   1 &  -0.1000D+01 &  -0.5429D+00 &   0.5429 &   0.1000D+01 &   0.5395D+00 &   0.5395\\
   2 &  -0.3464D+01 &  -0.3129D+01 &   0.9032 &   0.4641D+00 &   0.3927D+00 &   0.8461\\
   3 &  -0.7899D+01 &  -0.7714D+01 &   0.9766 &   0.1530D+00 &   0.1375D+00 &   0.8988\\
   4 &  -0.1442D+02 &  -0.1430D+02 &   0.9915 &   0.4137D-01 &   0.3814D-01 &   0.9220\\
   5 &  -0.2299D+02 &  -0.2289D+02 &   0.9956 &   0.1004D-01 &   0.9408D-02 &   0.9368\\
   6 &  -0.3357D+02 &  -0.3347D+02 &   0.9971 &   0.2282D-02 &   0.2161D-02 &   0.9470\\
   7 &  -0.4615D+02 &  -0.4606D+02 &   0.9980 &   0.4959D-03 &   0.4733D-03 &   0.9544\\
   8 &  -0.6074D+02 &  -0.6064D+02 &   0.9985 &   0.1043D-03 &   0.1002D-03 &   0.9600\\
   9 &  -0.7732D+02 &  -0.7723D+02 &   0.9988 &   0.2142D-04 &   0.2066D-04 &   0.9644\\
  10 &  -0.9591D+02 &  -0.9581D+02 &   0.9991 &   0.4314D-05 &   0.4176D-05 &   0.9680\\
  11 &  -0.1165D+03 &  -0.1164D+03 &   0.9992 &   0.8555D-06 &   0.8305D-06 &   0.9708\\
  12 &  -0.1391D+03 &  -0.1390D+03 &   0.9994 &   0.1675D-06 &   0.1630D-06 &   0.9733\\
  13 &  -0.1637D+03 &  -0.1636D+03 &   0.9995 &   0.3244D-07 &   0.3164D-07 &   0.9753\\
  14 &  -0.1902D+03 &  -0.1902D+03 &   0.9995 &   0.6228D-08 &   0.6084D-08 &   0.9770\\
  15 &  -0.2188D+03 &  -0.2187D+03 &   0.9996 &   0.1186D-08 &   0.1161D-08 &   0.9787\\
  16 &  -0.2494D+03 &  -0.2493D+03 &   0.9996 &   0.2247D-09 &   0.2198D-09 &   0.9784\\
  17 &  -0.2820D+03 &  -0.2819D+03 &   0.9997 &   0.4206D-10 &   0.4139D-10 &   0.9839\\
\hline
\end{tabular}
\end{table}
                                                                       
\end{document}